\documentclass[a4paper,11pt]{article}
\pdfoutput=1 
\usepackage{jheppub} 

\usepackage[T1]{fontenc} 

\pdfoutput=1
\usepackage{amsmath,amssymb}
\usepackage{amsfonts}
\usepackage{bm,bbm}
\usepackage{graphicx}
\usepackage{mathrsfs}
\usepackage{slashed}
\usepackage{booktabs}
\usepackage{tabu}
\usepackage[dvipsnames]{xcolor}
\usepackage[normalem]{ulem}
\usepackage{tikz}
\usepackage{soul}

\usepackage{hyperref}
\hypersetup{colorlinks,citecolor= blue,linkcolor= blue, urlcolor=blue}

\usepackage{xcolor}
\usepackage{ulem}
\usepackage{array}
\usepackage{verbatim}
\usepackage{epsfig}
\usepackage{multirow}

\newcommand{\be}{\begin{equation}}
\newcommand{\ee}{\end{equation}}
\newcommand{\ba}{\begin{array}}
\newcommand{\ea}{\end{array}}
\newcommand{\bea}{\begin{eqnarray}}
\newcommand{\eea}{\end{eqnarray}}

\usepackage{ulem,fancyvrb}
\usepackage{xcolor}

\title{Probing invisible dark photon models via atmospheric collisions}

\author[a,b]{Mingxuan Du,}
\author[a]{Rundong Fang,}
\author[a]{Zuowei Liu,}
\author[a]{Wenxi Lu,}
\author[a]{Zicheng Ye}

\affiliation[a]{Department of Physics, Nanjing University, Nanjing 210093, China}
\affiliation[b]{Center for High Energy Physics, Peking University, Beijing 100871, China}

\emailAdd{mingxuandu@pku.edu.cn}
\emailAdd{zuoweiliu@nju.edu.cn}
\emailAdd{luwenxi@smail.nju.edu.cn}
\emailAdd{zichengye@smail.nju.edu.cn}

\abstract{

Atmospheric collisions can copiously produce dark sector particles in the invisible dark photon model, leading to detectable signals in underground neutrino detectors. We consider the dark photon model with the mass mixing mechanism and use the Super-K detector to detect the electron recoil events caused by the atmospherically produced dark sector particles within the model. We find that the combined data from four Super-K runs yield new leading constraints for the invisible dark photon in the mass range of $\sim(0.5-1.4)$ GeV, surpassing various previous constraints, including those from BaBar and NA64.

}

\begin{document}

\maketitle
\flushbottom

\section{Introduction}

The nature of dark matter remains elusive, in spite of decades of dedicated efforts 
\cite{Bertone:2016nfn,Arbey:2021gdg}. In recent years there has been increased interest 
in models where dark matter resides in a dark sector and interacts with the standard model sector 
through the interactions of a dark photon \cite{Alexander:2016aln,Battaglieri:2017aum,Jaeckel:2010ni,Fabbrichesi:2020wbt}. 
Dark photon can arise in models where a new abelian gauge boson in the dark sector 
mixes with the hypercharge boson in the standard model 
via either the kinetic mixing mechanism \cite{Holdom:1985ag,Foot:1991kb} 
or the mass mixing mechanism \cite{Kors:2005uz,Feldman:2006ce,Feldman:2006wb,Feldman:2007wj,Du:2019mlc}.

Accelerator experiments are optimal environments  
for probing dark photons with masses above the MeV scale 
\cite{Fabbrichesi:2020wbt}. 
For dark photons with a substantial decay width into 
Standard Model (SM) particles, one of the key detection channels   
involves identifying distinct peaks 
in the invariant mass distribution of the decay products.
However, detecting dark photons that decay 
predominantly into dark matter final states is  
more challenging at accelerator experiments.
We refer to this kind of dark photon 
as the ``invisible dark photon.'' 
At the sub-GeV mass range, the current strongest constraints 
on the invisible dark photon come from 
the electron beam-dump experiment NA64 \cite{Banerjee:2019pds, Andreev:2023uwc} 
and the electron collider BaBar \cite{BaBar:2017tiz}. 
The corresponding detection channels are the missing energy signature
at NA64 \cite{Banerjee:2019pds, Andreev:2023uwc} 
and the mono-photon signature at BaBar \cite{BaBar:2017tiz}, 
respectively.

\begin{figure}[htbp]
\begin{centering} 
\includegraphics[width= 0.5\columnwidth]{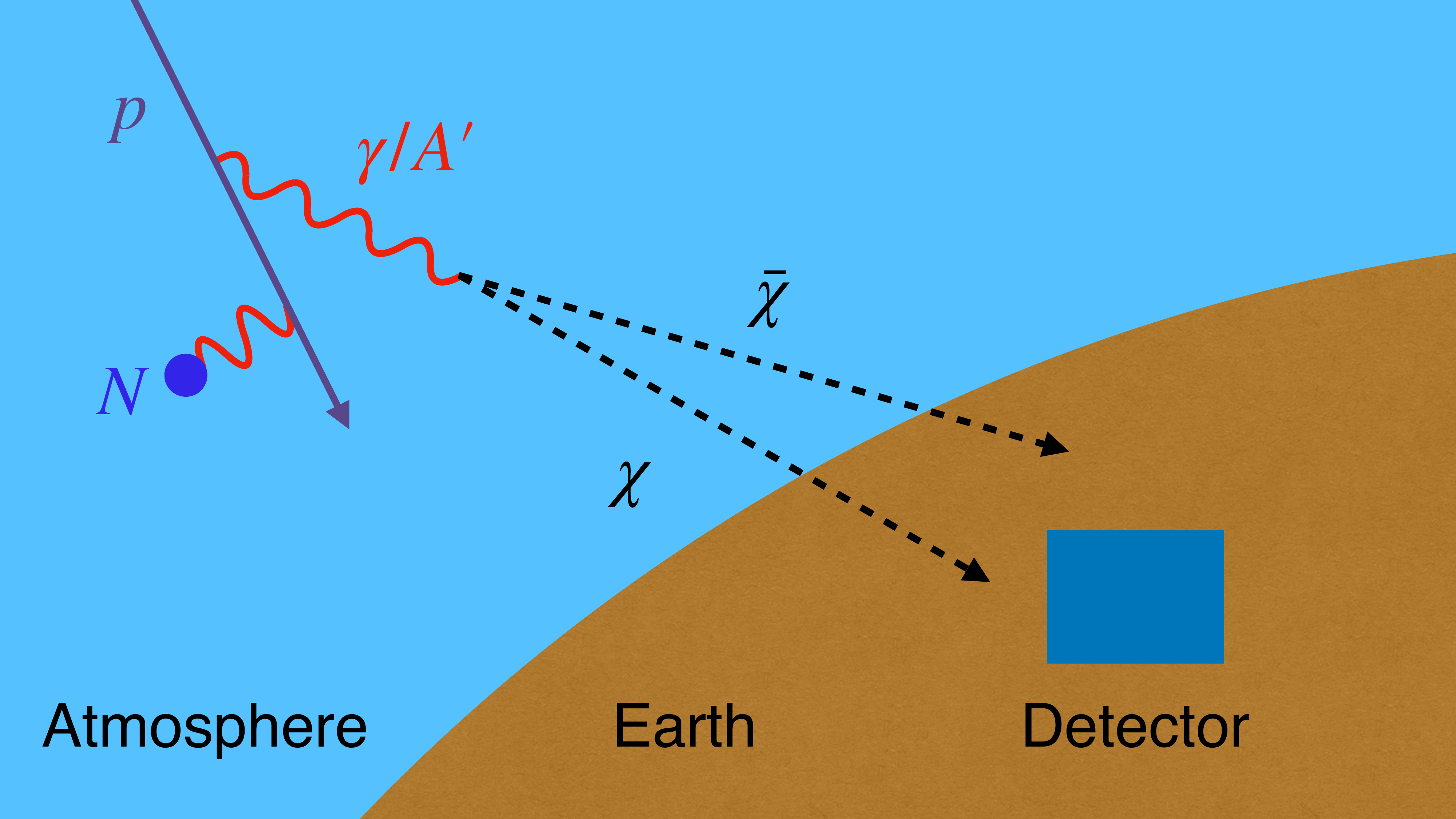}
\caption{Underground experiments detect dark sector particles  
generated in atmospheric collisions.}
\label{fig:pb-sketch}
\end{centering}
\end{figure}

In this study we consider 
cosmic ray interactions with the Earth's atmosphere  
as the source of dark matter production within the 
invisible dark photon model, 
as sketched in Fig.~(\ref{fig:pb-sketch}).
Recent studies have shown that the atmospheric collisions of  
cosmic rays 
can lead to a copious generation of sub-GeV dark matter particles, 
consequently presenting a novel avenue for probing the dark sector 
\cite{Alvey:2019zaa,Plestid:2020kdm,Su:2020zny,Kachelriess:2021man,ArguellesDelgado:2021lek,Iguro:2021xsu,Arguelles:2022fqq,Du:2022hms}. 
In this study, we consider the 
dark photon model with a mass mixing term 
and use the neutrino experiment Super-K 
to detect dark matter originating from atmospheric collisions. 
We find that the combined data from four Super-K runs 
yield new leading constraints 
on the invisible dark photon 
in the mass range of $\sim(0.5-1.4)$ GeV.

\section{The Model}
\label{sec:the-model}

We consider the dark photon model that has a mass mixing term: 
\begin{equation}
\mathcal{L}=\mathcal{L}_{\rm SM}-\frac{1}{4}  
\tilde F_{\mu \nu}^{\prime} 
\tilde F^{\prime \mu \nu} + 
\frac{\tilde m_{A^{\prime}}^2}{2} \left(\tilde A_\mu^{\prime} 
+ \epsilon \tilde A_{\mu}\right)^2 +\bar{\chi} (i\gamma^\mu \partial_\mu -m_\chi 
+g_d  \gamma^\mu \tilde A_\mu^{\prime} )\chi 
\label{eq:L}
\end{equation}
where 
$\tilde A$ is the SM photon, 
$\tilde A^{\prime}$ is the $U(1)_d$
gauge bosons in the dark sector, 
$\tilde F'_{\mu \nu}=
\partial_\mu \tilde A^{\prime}_{\nu}
-\partial_\nu \tilde A^{\prime}_{\mu}$, 
$\chi$ is the Dirac fermion in the dark sector 
charged under the $U(1)_d$ gauge, 
$m_\chi$ is the mass of $\chi$, 
$\tilde m_{A^{\prime}}$ is the gauge boson mass term, 
and $\epsilon$ is the dimensionless mass mixing parameter.
The gauge boson mass terms 
in Eq.~\eqref{eq:L} 
can be naturally generated in the Stueckelberg mechanism; 
see e.g., Ref.~\cite{Feldman:2007wj} for 
such 
mass terms for the hypercharge gauge boson in the SM sector 
and the $U(1)_d$ gauge boson in the dark sector.

The mass matrix can be diagonalized by the following transformation 
\cite{Feldman:2007wj} 
\begin{equation}
	\begin{bmatrix}
		\tilde A_{\mu}\\
		\tilde A^{\prime}_{\mu}
	\end{bmatrix}
 =\kappa
\begin{bmatrix}
1&\epsilon \\
-\epsilon&1
\end{bmatrix}
\begin{bmatrix}
		A_{\mu}\\
		A^{\prime}_{\mu}
	\end{bmatrix},
 \label{eq:trans}
\end{equation}
where $\kappa \equiv 1/\sqrt{1+\epsilon^2}$. 
Note that here we consider only the mass terms for 
the two gauge bosons in Eq.~\eqref{eq:L}, 
and neglect the neutral gauge boson mass terms 
that arise from the Higgs mechanism in the SM sector.
The mass diagonalization in Eq.~\eqref{eq:trans}  
leads to a massless state $A_{\mu}$, 
which can be identified as the 
SM photon (the mass eigenstate),
and a massive eigenstate $A^\prime_{\mu}$ 
with a mass of 
$m_{A'} = \tilde m_{A'} /\kappa$, 
which is the dark photon. 
In the parameter space of interest 
in this analysis, one has $\kappa \simeq 1$. 
The interaction Lagrangian in terms of 
the mass eigenstates is 
\begin{equation}
\mathcal{L}_{\rm int}= \kappa 
(\epsilon J^{\rm EM}_{\mu}+J'_{\mu})
A^{\prime \mu}+
\kappa (J^{\rm EM}_{\mu}-\epsilon J'_{\mu}) A^{\mu} , 
\label{eq:L1}
\end{equation}
where $J^{\rm EM}_\mu$ and $J_\mu^{\prime}=g_d\bar{\chi}\gamma_{\mu}\chi$ 
are the currents in the standard model sector 
and the dark sector, respectively. 
The coupling between $\chi$ and $A_{\mu}$ 
is suppressed by the small parameter $\epsilon$, 
leading to a millicharge 
of $Q_\chi = g_d\epsilon \kappa/e$ for $\chi$, 
where $e$ is the QED coupling. 
Note that in this model, 
the dark sector fermion $\chi$ 
interacts with both the dark photon and the photon. 
This is different from the kinetic-mixing scenario, 
where the interaction between the SM sector and 
the dark sector is mediated via 
$(\epsilon/2) \tilde F'_{\mu\nu} \tilde F^{\mu\nu}$ 
with $\epsilon$ being the kinetic-mixing parameter. 
In such models, 
a massive $A'$ and a millicharged $\chi$ 
cannot appear simultaneously \cite{Feldman:2007wj}. 
Consequently, photon-mediated processes are absent,
leading to weaker bounds  
in the kinetic mixing case,  
as shown in Fig.~(\ref{fig:bg-agnostic}).

In this study, we consider the following 
parameters:
$\epsilon \ll 1$, 
$\alpha_d \equiv g_d^2 /4\pi = 0.1$,
and $m_{A'} = 3 m_\chi$; 
in this case,
the dark photon decays predominately into $\chi$ 
and its decay width into 
the standard model particles is negligible. 
Thus, we use the invisible decay width 
$\Gamma(A'\to \chi \bar\chi$) to approximate the total 
decay width: 
\be
\Gamma_{A'} = \frac{\alpha_d}{3} m_{A'} \left(1 + \frac{2 m^2_{\chi}}{m^2_{A'}}\right)\sqrt{1 - \frac{4 m^2_{\chi}}{m^2_{A'}}}. 
\label{eq:decay_width}
\ee

\section{Atmospheric flux}
\label{sec:atmospheric-flux}

In this section we compute the flux of $\chi$
generated by cosmic ray collisions with the atmosphere, 
by using the one-dimension approximation \cite{Gondolo:1995fq}. 
The flux of $\chi$ on the Earth surface is isotropic 
in the one-dimension approximation and is given by
\cite{Gondolo:1995fq} \cite{Du:2022hms}
\be
\frac{d^2\Phi^s_\chi}{dE^s_\chi d\Omega^s_\chi}=
\iint dhdE_p\frac{d^2\Phi_p(h)}{dE_pd\Omega_p}
n_T(h)\sigma_{pT}\sum_i\frac{dN^i_\chi}{dE^s_\chi},
\label{eq:mcpflux}
\ee
where 
$h$ is the altitude, 
$E^s_\chi$ ($E_p$) is the energy of $\chi$ (proton),  
$\Omega^s_\chi$ ($\Omega_p$) is the solid angle of $\chi$ (proton), 
$\Phi^s_\chi$ is the flux of $\chi$ at the surface of Earth, 
$\Phi_p(h)$ is the cosmic ray proton flux at altitude $h$, 
$n_T(h)$ is the number density of the target nucleus $T$
in the atmosphere at altitude $h$, 
$\sigma_{pT}$ is the inelastic proton-air cross section, 
and ${dN^i_\chi}/{dE^s_\chi}$ is the energy spectrum of $\chi$ 
per proton-air collision in the production channel 
denoted by the superscript $i$. 
We use CRMC \cite{Pierog:2013ria, ulrich_ralf_2021_4558706} to compute $\sigma_{pT}$ for nitrogen and oxygen; 
because of the weak energy dependence in the region of interest, 
we use a constant cross section 
$\sigma_{pT}\simeq $ 253 (281) mb 
for nitrogen (oxygen). We use 
the NRLMSISE-00 atmosphere model~\cite{picone2002nrlmsise}
for the air density $n_T(h)$. 
To compute the cosmic ray proton flux at altitude $h$, 
we solve
the cascade equation~\cite{Gondolo:1995fq} 
\begin{equation}
	\frac{d}{dh}
 \frac{d^2\Phi_p(h)}{dE_pd\Omega_p}=\sigma_{pT}n_T(h)\frac{d^2\Phi_p(h)}{dE_pd\Omega_p}, 
\end{equation}
with the power law distribution for cosmic ray protons 
at $h_{\rm max}=65$ km
\cite{ParticleDataGroup:2018ovx} 
\begin{equation}
	\frac{d^2\Phi_p(h_{\rm max})}{dE_pd\Omega_p}=\frac{0.74\times1.8\times10^4}
 {\text{m}^2 \, \text{s} \, {\rm sr} \, {\rm GeV}} 
 \left( \frac{E_p}{\rm GeV} \right)^{-2.7}. 
\label{eq:power-law}
\end{equation}

There are two production channels of $\chi$ in atmospheric 
collisions: 
production of $\chi$ in meson decays 
{\cite{Plestid:2020kdm,Kachelriess:2021man,ArguellesDelgado:2021lek}, 
and production of $\chi$ in the proton bremsstrahlung process \cite{Du:2022hms}.
As recently pointed out by Ref.~\cite{Du:2022hms}, 
the proton bremsstrahlung process is the dominant 
process of producing millicharged particles 
in the sub-GeV mass range. 
For example, 
the millicharged particle flux arising from meson decays is 
less than $0.5\%$ of that from the proton bremsstrahlung process for mass $\sim 0.2$ GeV \cite{Du:2022hms}. 
Since the dark sector fermion $\chi$ in the model considered here is similar to 
millicharged particles in the minimal model  
considered in Ref.~\cite{Du:2022hms}, 
we only consider the proton bremsstrahlung process 
as the production channel.

\begin{figure}[t]
\centering
\includegraphics[width=0.4 \textwidth]{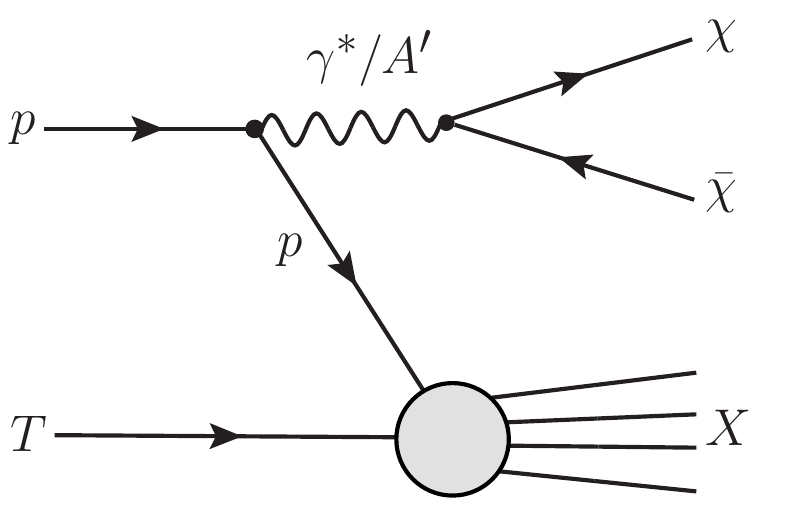} 
\caption{Production of $\chi$ in the proton bremsstrahlung process 
in collisions between the cosmic proton $p$ and the nucleus $T$
in the atmosphere.}
\label{fig:PB}
\end{figure}

The Feynman diagrams of the proton bremsstrahlung 
process for the dark photon model considered in this study 
are shown in Fig.~(\ref{fig:PB}); 
the mediator to produce a pair of $\chi$'s can be either a  photon or a dark photon.
Ref.~\cite{Du:2022hms} computed the production of 
millicharged particles 
in the proton bremsstrahlung process 
where only the virtual photon mediator was taken into account. 
Here we further include the contribution from the 
dark photon mediator. 
Thus, the energy spectrum of $\chi$ in the 
proton bremsstrahlung process is given by 
\bea
     \frac{d N_\chi}
     {d E_\chi} = 
  \frac{\epsilon^2 g_d^2}{6 \pi^2 }
\int \frac{d k^2}{k^2}\frac{(2k^2-m^2_{A'})^2 + m^2_{A'} \Gamma^2_{A'}}{(k^2-m^2_{A'})^2 + m^2_{A'} \Gamma^2_{A'}}
\sqrt{1 - 4x}
\nonumber \\ \times 
\left( 1 + 2x \right) 
\int 
\frac{d E_k}{\sigma_{pT}}
\frac{d \sigma_{\rm PB}}{d E_k} 
\frac{\Theta\left(E_\chi- E_{-}\right) \Theta\left(E_{+} - E_\chi \right)}{E_{+} - E_{-}}, 
\label{eq:PB:MCP:spectra}
\eea
where 
$E_\chi$ is the energy of $\chi$ in the lab frame, 
$k^\mu=(E_k,\vec k)$ is the momentum of the virtual photon 
$\gamma^*$ or the dark photon $A'$, 
$\Gamma_{A'}$ is the total decay width of the dark photon, 
$x={m_\chi^2}/{k^2}$, 
${d \sigma_{\rm PB}}/{d E_{k}}$ 
is the differential cross section 
for the inclusive process of $pN\to \gamma^* X$, 
$E_{\pm} \equiv \gamma (E_{\chi}^r 
\pm  \beta p_{\chi}^r)$ denote  
the maximal and minimum values of $E_\chi$
with 
$\gamma = (1-\beta^2)^{-1/2}
= E_{k}/\sqrt{k^2}$, 
and $E_\chi^r$ ($p_\chi^r$) is 
the energy (magnitude of momentum) of 
$\chi$ in the rest frame of $\gamma^*/A'$.

In our analysis, we adopt the expression for ${d \sigma_{\rm PB}}/{d E_{k}}$ 
obtained in Ref.~\cite{Du:2022hms}, which incorporates the form factor  
\begin{equation}
F_{*}=\frac{\Lambda^4}{\Lambda^4+(p^{\prime 2}-m_p^2)^2} 
\label{eq:FF}
\end{equation}
to account for the off-shell nature of the intermediate proton 
\cite{Foroughi-Abari:2021zbm}, 
where $p^{\prime}$ denotes the four-momentum of the intermediate proton 
(as shown in Fig.~\ref{fig:PB}) 
and $\Lambda$ is the scale parameter. 
To constrain $\Lambda$,  
Ref.~\cite{Du:2022hms} analyzed di-muon production data 
from the NA60 experiment \cite{NA60:2016nad}, 
which receive contributions from the proton bremsstrahlung process, 
as shown in Fig~\ref{fig:dimuon}. 
The NA60 experiment is a fixed-target experiment  
with an angular acceptance between 35 and 120 mrad. 
In Ref.~\cite{Du:2022hms}, 
the detection probability for the di-muon final state
was approximated by the case where the virtual photon 
has a polar angle of $\theta_k = 77.5$ mrads. 
In the current study, we do not adopt such an approximation;  
instead, we explicitly require both $\mu^+$ and $\mu^-$ 
to fall within the angular range of $35-120$ mrad.

\begin{figure}[t]
\centering
\includegraphics[width=0.4 \textwidth]{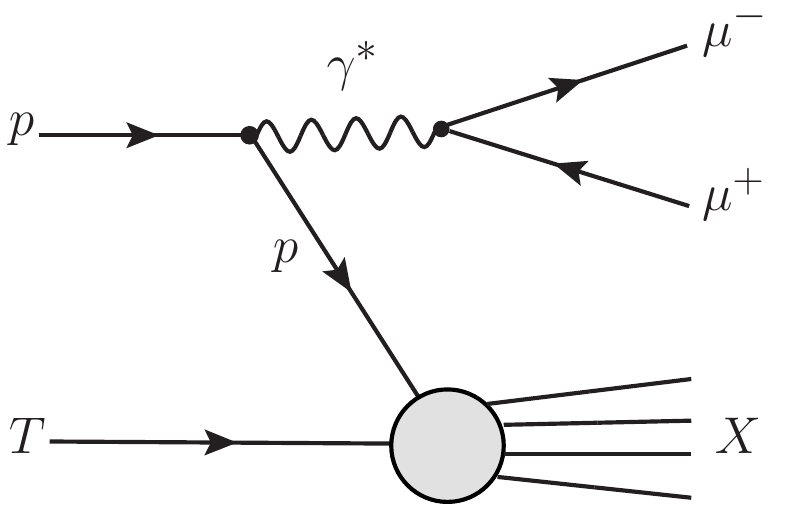} 
\caption{Feynman diagram for di-muon production 
in the proton bremsstrahlung process  
in collisions between the incident proton and the NA60 target.}
\label{fig:dimuon}
\end{figure}

\begin{figure}[htbp]
\centering
\includegraphics[width=0.5 \textwidth]{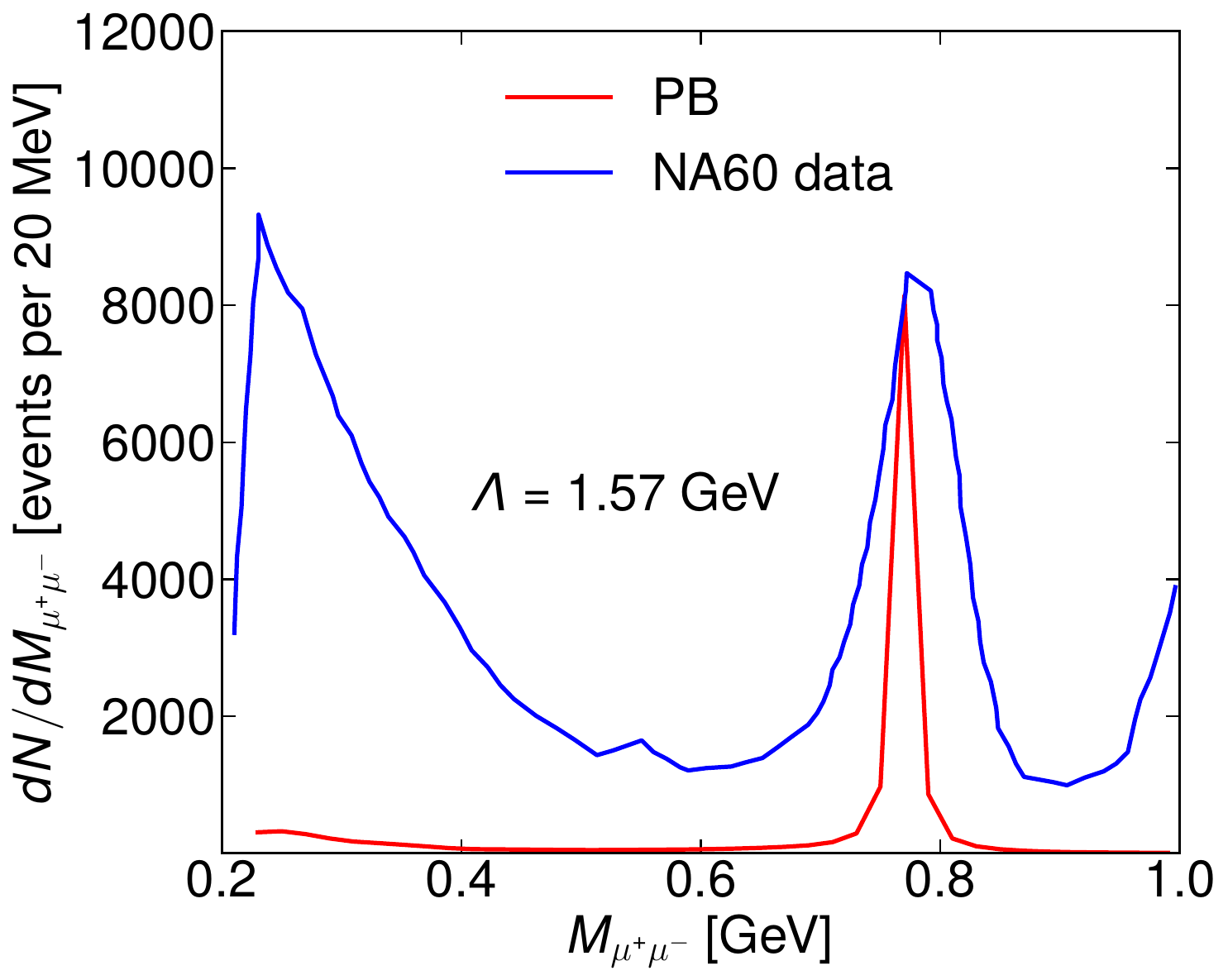} 
\caption{Di-muon events as a function of the invariant mass in the NA60 experiment: 
(1) NA60 data from Ref.~\cite{NA60:2016nad} (blue), 
(2) predictions from the proton bremsstrahlung process (red).}
\label{fig:di-muon-2}
\end{figure}

Fig.~\ref{fig:di-muon-2} compares the di-muon mass spectrum 
predicted by the proton bremsstrahlung process with the NA60 measurements, 
where the data are binned with a bin size of 20 MeV. 
The event rate from the proton bremsstrahlung process 
decreases with decreasing $\Lambda$ in Eq.~\eqref{eq:FF}. 
To derive a bound on $\Lambda$, we require 
the prediction from the proton bremsstrahlung process 
not to exceed the NA60 data in any bin. 
The most stringent constraint is provided by the $0.76$–$0.78$ GeV bin. 
By saturating the prediction from the proton bremsstrahlung process 
in this bin, 
we obtain $\Lambda = 1.57$ GeV. 
However, since meson decays also contribute to 
the dimuon spectrum, 
the contribution from the proton bremsstrahlung process 
should be correspondingly reduced. 
Therefore, $\Lambda=1.57$ GeV should be 
regarded as an upper bound on $\Lambda$.  
We also note that 
there are possible double countings  
between the proton bremsstrahlung process and the meson 
decay process, which are beyond the scope of the current analysis.

\section{Earth attenuation}
\label{sec:earth-attenuation}

In order to precisely obtain the flux of $\chi$ 
at underground neutrino detectors, 
it is necessary to take into account the Earth's attenuation effect. 
Charged particles lose energy 
through ionization and radiative processes 
as they traverse a medium \cite{ParticleDataGroup:2022pth};  
for low-energy charged particles, 
ionization is usually the dominant process. 
Since the mass of $\chi$ of interest in this analysis
falls within the range of $\sim (0.1-1)$ GeV, we follow Ref.~\cite{ArguellesDelgado:2021lek} and use Earth's attenuation effects on muons 
as a reference to account for the energy loss of $\chi$.

Thus, the energy loss of $\chi$ along the trajectory traversed in Earth 
can be computed as follows 
\cite{Gaisser:2016uoy, ArguellesDelgado:2021lek,Hu:2016xas}
\be
- \frac{d E}{d X} =  
Q_{\chi}^2 (a + b E),
\label{eq: energy loss1}
\ee
where $X$ is the slant depth traversed, 
and $a$ ($b$) is the parameter to describe energy loss due to 
ionization (radiation in scatterings 
with nuclei).  
In our analysis we adopt the parameters for muons in the standard rock: 
$a=0.233\ \rm GeV/mwe$ and $b=4.64\times 10^{-4}\ \rm mwe^{-1}$ 
where $1\ \rm mwe=100\ g/cm^{2}$ \cite{Koehne:2013gpa}.
The slant depth $X$ for the Super-K detector is given by $X=\rho L$
where $\rho=2.6\ \rm g/cm^{3}$ is the mass density of the standard rock, 
and $L$ is the distance travelled in the rock. 
We compute $L$ via 
\begin{equation}
   L=\sqrt{(2 R_e^2 - 2 R_e d)
   \left(1-\cos
   \left(\theta-\theta_s\right) \right) + d^2},
\end{equation}
where
$R_e$ is the radius of Earth, 
$d=1\ \rm km$ is the depth of the Super-K detector 
\cite{Super-Kamiokande:2002weg}, 
$\theta_s$ is the zenith angle of the MCP at the Earth surface, 
and $\theta$ is the zenith angle of the Super-K detector. 
By solving Eq.~\eqref{eq: energy loss1}, the MCP flux at the Super-K detector is given by 
\be
\frac{d^2 \Phi_\chi^D \left(X\right)}{dE_\chi^D d \Omega^D}
=\textrm{exp} \left(Q_{\chi}^2bX\right)
\frac{d^2\Phi_\chi^s}{dE_\chi^sd\Omega^s},
\ee
where the superscripts $D$  and $s$ denote the 
physical quantities at the detector 
and at the Earth surface, respectively, and $E_\chi^s=\left(E_\chi+a/b\right)\ \textrm{exp} \left(Q_{\chi}^2 bX\right)-a/b$.

\section{Signals at Super-K}
\label{sec:Signals-at-Super-K}

The Super-K experiment, 
a large water-Cherenkov detector 
with a fiducial volume of 22.5 kton of water
\cite{Super-Kamiokande:2002weg}, 
is the ideal place to search for dark sector 
particles produced in atmospheric collisions 
\cite{Plestid:2020kdm,ArguellesDelgado:2021lek,Du:2022hms,Arguelles:2022fqq,Iguro:2021xsu} 
and in Earth-bound dark matter annihilations 
\cite{McKeen:2023ztq}.
For the atmospherically generated dark fermion $\chi$ in
the invisible dark photon model,
the dominant signals in Super-K come from
the elastic scattering between 
$\chi$ and the target electron, 
mediated by either the 
photon or the dark photon. 
Thus, the $\chi-e$ scattering cross section is given by 
\be
\begin{aligned}
\frac{d \sigma}{\mathrm{d} E_r} =&  \pi \alpha 
{\alpha_d} \epsilon^2\frac{E_{r}+2 E_{\chi}^{2}/E_{r} - 
2 E_{\chi} - m_{e} - m_{\chi}^{2}/m_e}{(E^2_{\chi}-m^2_{\chi})m_eE_r}\\& 
\times \left(\frac{4m_eE_r+m_{A^\prime}^2}{2m_eE_r+m_{A^\prime}^2}\right)^2,    
\end{aligned}
\label{eq:xec-dp-electron}
\ee
where 
$E_\chi$ is the energy of $\chi$,
$E_r$ is the recoil energy of the electron, 
and $m_e$ is the electron mass.

In our analysis we use electron recoil
data from four Super-K runs with a total exposure of 176 kton-year 
\cite{Super-Kamiokande:2011lwo, Super-Kamiokande:2021jaq}.  
The electron recoil data in Super-K are selected by 
the Cherenkov angle in the range of 38-50 degrees, 
since relativistic charged particles 
(such as electrons in the energy range of interest) 
have a Cherenkov angle of 42 degrees, 
whereas heavier particles have a smaller Cherenkov angle \cite{Super-Kamiokande:2011lwo, Super-Kamiokande:2021jaq}. 
The 
Super-K I-\uppercase\expandafter{\romannumeral3} data  
in the electron recoil energy range of 16-88 MeV 
are binned into 18 bins with 
a bin width of 4 MeV for each data bin 
\cite{Super-Kamiokande:2011lwo}; 
the Super-K \uppercase\expandafter{\romannumeral4} data 
in the electron recoil energy range of 16-78 MeV 
are binned into 31 bins with 
a bin width of 2 MeV for each data bin 
\cite{Super-Kamiokande:2021jaq}. 
We compute the signal events $S_i$ in each data bin via 
\be
S_i=2\pi n_e \mathcal{E} \int ~dE_r f(E_r)\int~dE_\chi ~dz\frac{d^2\Phi^D_\chi}{dE_\chi~dz}\frac{~d\sigma}{~dE_r},
\ee
where $z\equiv \cos\theta$, 
$n_e$ is the number of electrons per ton of water, 
$\mathcal{E}$ is the exposure, 
and $f(E_r)$ is the detector efficiency. 
In our analysis we adopt the efficiency curve in Fig. (10) of Ref.~\cite{Super-Kamiokande:2011lwo}.

\section{Angular distributions}
\label{sec:Angular-distributions}

Because of Earth's attenuation, in the parameter space of interest, 
the $\chi$ flux with $\theta > 90^\circ$ are significantly suppressed 
compared to that with $\theta < 90^\circ$,
where $\theta$ is the zenith angle. 
However, the background events 
are not suppressed in the $\theta > 90^\circ$ region. 
Note that, in the parameter space of interest,
the recoiled electron has almost the
same direction as the incident $\chi$.
Thus, to suppress background events, we  
take the electron recoil data 
in the range of $\theta < 90^\circ$ 
as the signal region.

In the low energy region of the electron recoil data 
(16 MeV $ \lesssim E_r \lesssim$ 55 MeV), 
the dominant backgrounds are due to  
electrons from decays of very low-energy muons 
that originate in the charged-current (CC) process of 
muon-neutrinos 
\cite{Super-Kamiokande:2011lwo, Super-Kamiokande:2021jaq}. 
These low-energy muon decays give rise to  
an isotropic electrons background. 
In the high energy region 
(55 MeV $\lesssim E_r \lesssim$ 88 MeV), 
the dominant backgrounds come from the CC process
of the electron-neutrinos 
\cite{Super-Kamiokande:2011lwo, Super-Kamiokande:2021jaq}. 
The angular distribution of the 
electron-neutrino flux is not isotropic: 
it has a peak near $\theta = 90^\circ$ 
and exhibits an upward-downward asymmetry with 
more neutrinos 
from the region of $\theta > 90^\circ$ 
compared to the region of $\theta < 90^\circ$
\cite{Taani:2020rrm}. 
For example,  
the electron-neutrino flux with energy of $0.32$ GeV
in the region of $\theta < 60^\circ$ 
is about 18\%  
of the flux from all angles
\cite{Taani:2020rrm}. 
However, 
the $\chi$ flux in the region of $\theta < 60^\circ$ 
is about 50\%
of the signal flux from all angles.
Thus, we multiply a factor of half to 
the data events in Ref.~\cite{Super-Kamiokande:2011lwo, Super-Kamiokande:2021jaq} 
to obtain the data events in the signal region, $\theta < 90^\circ$. 
We note that one could take advantage of the different 
angular distributions of the signal and background events 
in the high energy region to further improve the constraints.

\section{Results}
\label{sec:Results}

To constrain dark photon models,
we take a background-agnostic approach 
\cite{ArguellesDelgado:2021lek}
\cite{Arguelles:2019ouk}
such that the likelihood function is constructed only by the data bins 
in which the predicted signal events (without background) 
are more than the observed data events. 
For each data bin, 
we construct the likelihood function via 
\bea
\mathcal{L}_i = \begin{cases}P\left(D_i|S_i\right) & D_i \leq S_i \\ 1 & D_i>S_i\end{cases},
\eea
where $P\left(D_i|S_i\right)$ is the Poisson distribution: 
\be
P\left(D_i|S_i\right)=\frac{(S_i)^{D_i}}{D_i!}\mathrm{exp}(-S_i),
\ee
with $D_i$ and $S_i$ being 
the data events and expected signal events in the $i$-th bin, respectively. 
The total likelihood is given by 
${\cal L}= \Pi_i {\cal L}_i$, 
and the test statistics is given by 
\begin{equation}
    \mathcal{TS} = - 2 \log \left[ 
    \frac{{\cal L}(m_\chi, \varepsilon)}{{\cal L}(m_\chi,\varepsilon=0)} 
    \right].
\end{equation}
We use $\mathcal{TS} < 4.6$ to 
set the $90\%$ confidence level upper bound on the dark photon model.

Fig.~(\ref{fig:bg-agnostic}) shows the Super-K 
$90\%$ CL constraints on the mass mixing parameter $\epsilon$ in
the invisible dark photon model, 
where we use the data from 
four Super-K runs with a total exposure of 176 kton-year. 
For the invisible dark photon 
in the mass range of $\sim(0.1-1.4)$ GeV 
(except the small region near $\sim 0.2$ GeV), 
the Super-K constraints surpass 
the current best limits from 
NA64 \cite{Andreev:2023uwc,Andreev:2023qsw} and BaBar \cite{BaBar:2017tiz}.

Because in our dark photon model, 
the dark sector fermion $\chi$ is millicharged, 
this model also receives constraints from searches of 
millicharged particles. 
Fig.~(\ref{fig:bg-agnostic}) shows 
the currently best millicharge constraints in the parameter 
space of interest, including 
SLAC mQ \cite{Prinz:1998ua}, 
LSND \cite{Magill:2018tbb}, 
BEBC \cite{Marocco:2020dqu}, 
and SENSEI \cite{SENSEI:2023gie}. 
As shown in Fig.~(\ref{fig:bg-agnostic}), 
the Super-K limits analysed in this work are 
better than all other constraints 
for dark photon 
in the mass of $\sim(0.69-0.87)$ GeV.

\begin{figure}[t]
\centering
\includegraphics[width=0.5\textwidth]{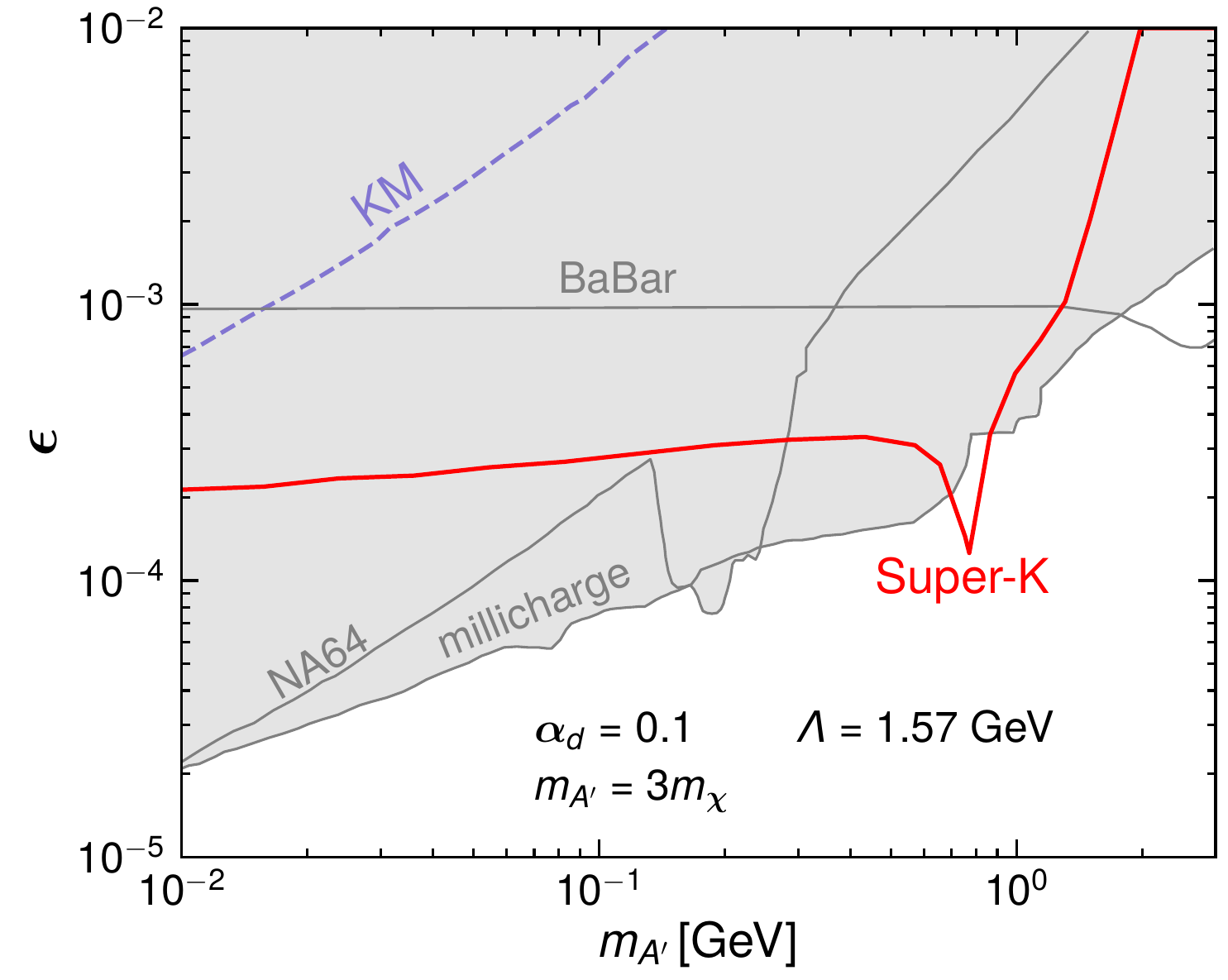} 
\caption{Super-K constraints ($90\%$ CL) on the mass mixing parameter $\epsilon$ 
in the invisible dark photon model with $\Lambda=1.57$ GeV (red solid). 
Super-K constraints on the 
kinetic mixing dark photon model 
are also shown (blue dashed). 
Existing constraints are shown as gray shaded regions:  
constraints on the invisible dark photon model 
from 
NA64 \cite{Andreev:2023uwc,Andreev:2023qsw} 
and 
BaBar \cite{BaBar:2017tiz}, 
and constraints on the millicharge of the dark fermion.
For the dark photon model, 
we fix $m_{A'} = 3 m_{\chi}$ and $\alpha_d = 0.1$.}
\label{fig:bg-agnostic}
\end{figure}

The results shown in 
Fig.~(\ref{fig:bg-agnostic}) 
are obtained 
with the power-law spectrum of cosmic protons given in 
Eq.~\eqref{eq:power-law}. 
However, low-energy protons are prevented 
by the geomagnetic field from scattering with the main distribution 
of the atmosphere at low altitude. 
Ref.~\cite{Du:2022hms} found that the results obtained with the 
power-law spectrum are consistent with 
the actual cosmic proton data, such as those from AMS-02 \cite{AMS:2015tnn}. 
To estimate the effects of the geomagnetic field, 
we first equate the altitude of the AMS-02 experiment  
with the gyroradius of the proton at the Super-K 
to determine the cut-off energy on the cosmic proton, 
which is found to be $\sim 5$ GeV. 
We then recompute the Super-K limits by excluding cosmic 
protons with $E<5$ GeV in Eq.~\eqref{eq:power-law}, 
and find that the Super-K limits shown in 
Fig.~(\ref{fig:bg-agnostic})  
are unchanged in the dark photon mass range of $\gtrsim 0.5$ GeV.  
The above statement holds even for a cut-off energy of 10 GeV. 
However, the geomagnetic field does have 
noticeable effects in the dark photon mass range of $\lesssim 0.5$ GeV, resulting in somewhat weaker limits.

The stronger constraints in the vicinity of the dark photon mass 
of $\sim$0.77 GeV are primarily due to the time-like form factor 
of the various $\rho/\omega$ vector mesons 
\cite{Du:2022hms,deNiverville:2016rqh,Feng:2017uoz,Foroughi-Abari:2021zbm,Du:2021cmt,Faessler:2009tn}, 
which are included in the computation of $d\sigma_{\rm PB}/dE_k$ 
in Eq.~\eqref{eq:PB:MCP:spectra}; 
see Ref.~\cite{Du:2022hms} for more details.
This is because when the dark photon mass is close to the 
lowest $\rho/\omega$ meson mass ($\sim 0.77$ GeV), 
there is an enhancement on the signal from 
the Breit-Wigner distributions  
both of the dark photon and 
of the time-like form factor of $\rho/\omega$.

We note that the constraints on millicharged particles shown in Fig.~(\ref{fig:bg-agnostic}) are 
analyzed in the minimal model where the effects of dark photon are neglected. 
The presence of a dark photon with mass comparable to the millicharged particle could 
potentially alter the signal significantly. 
We leave this to a future study.

\section{Conclusions}
\label{sec:Conclusions}

In this study, we study the Super-K 
constraints on the invisible dark photon model 
with a mass mixing parameter. 
We use cosmic proton collisions with the 
Earth's atmosphere to copiously produce the dark sector 
fermion in the invisible dark photon model, 
which then leads to a detectable electron recoil signal 
in the Super-K detector. 
By carefully investigating the Earth's attenuation effects and 
the angular distributions of both the signal events and the background events, 
we find that the combined data from the four Super-K runs provide   
a new leading constraint on the invisible dark photon  
in the mass of $\sim(0.69-0.87)$ GeV.

\begin{acknowledgments}
{\it Acknowledgments.---} 
We thank Li-Gang Xia and Yong-Heng Xu for discussions. 
The work is supported in part by the 
National Natural Science Foundation of China under Grant Nos.\ 12275128 
and 12147103.
\end{acknowledgments}

\bibliography{ref.bib}{}
\bibliographystyle{utphys28mod}

\end{document}